\documentstyle{article}

\newcommand{\ba}{\begin{eqnarray}}
\newcommand{\ea}{\end{eqnarray}}
\newcommand{\s}{\scriptsize}
\newcommand{\m}{\mbox}
\begin{document}
\def\appendix{\setcounter{section}{0}
\def\thesection{APPENDIX \Alph{section}:}
\def\theequation{\Alph{section} \arabic{equation}}}
\renewcommand{\thesection}{\Roman{section}.}
\renewcommand{\thesubsection}{\Alph{subsection}.}
\pagestyle{plain}
\def\ii{\'\i}
\begin{center}
{\bf\Large
Transformation brackets between
$U(\nu+1) \supset U(\nu) \supset SO(\nu)$ and
$U(\nu+1) \supset SO(\nu+1) \supset SO(\nu)$
}
\end{center}

\begin{quote}
Elena Santopinto\\
{\em Dipartimento di Fisica dell'Universit\`a di Genova,
 Genova, Italy and Istituto Nazionale di Fisica Nucleare,
Sezione di Genova,via Dodecaneso 33, 16164 Genova, Italy}
\end{quote}

\begin{quote}
Roelof Bijker\\
{\em Instituto de Ciencias Nucleares, U.N.A.M., A.P. 70-543,
 04510 M\'exico, D.F., M\'exico}
\end{quote}

\begin{quote}
Francesco Iachello\\
{\em Center for Theoretical Physics, Sloane Laboratory,
Yale University, New Haven, CT 06520-8120, USA}
\end{quote}

\begin{quote}
We derive a general expression for the transformation brackets
between the chains 
$U(\nu+1) \supset U(\nu) \supset SO(\nu)$ and 
$U(\nu+1) \supset SO(\nu+1) \supset SO(\nu)$ for $\nu \geq 2$.
\end{quote}
\begin{quote}
\end{quote}

\newpage
\section{INTRODUCTION}

The properties of bound states in a large variety of 
physical systems can be described by writing 
the Hamiltonian and other operators in terms of a spectrum 
generating algebra, $G$. In many applications the 
spectrum generating algebra is taken to be the unitary 
algebra $G=U(\nu+1)$ $^{\m{\s{1}}}$, where $\nu$ denotes 
the dimension. 
Examples of this approach are: the description of the
five quadrupole degrees of freedom of the interacting
boson model in nuclear physics in terms of the algebra
$U(6)$ $^{\m{\s{2}}}$, the description of the three dipole
degrees of freedom of the vibron model in molecular
physics in terms of the algebra $U(4)$ $^{\m{\s{3}}}$,
and the description of the six degrees of freedom (two
dipoles) of the valence quark model of baryons in
hadronic physics in terms of the algebra
$U(7)$ $^{\m{\s{4}}}$. The algebra $U(\nu+1)$ always admits
(for $\nu \geq 2$) two subalgebra chains
\begin{equation}\label{a1}
\begin{array}{ccccccc}
& &  U(\nu) & & & & \hspace{2cm} \m{(I)} \\
& \nearrow &  & \searrow & & & \\
 U(\nu + 1) & & & & SO(\nu) & , & \\
& \searrow &  & \nearrow & & & \\
& &  SO(\nu+1) & & & & \hspace{2cm} \m{(II)} \\
\end{array}
\end{equation}
in addition, eventually, to other chains. (The special case,
$\nu=1$, with $U(2) \supset U(1)$ and $U(2) \supset SO(2)$,
in which the two subalgebras $U(1)$ and $SO(2)$ are
isomorphic is treated in detail in Refs.~3 and~5 and it
will not be discussed further.)
In the applications mentioned above, the 
first chain has the physical meaning of a spherical
oscillator in $\nu$ dimensions with $U(\nu)$ being the degeneracy 
algebra, while the second chain has
the meaning of a displaced (or deformed) oscillator in
$\nu$ dimensions. 
The best known example of the latter
is the $SO(6)$ chain of the interacting boson model $^{\m{\s{6}}}$,
which has played an important role in nuclear structure physics.
       In view of the fact that the algebraic method is
presently being applied to a variety of problems in
physics with different number of dimensions it
is of interest to derive a general expression for the
transformation brackets between the two chains given in
the group lattice of Eq.~(\ref{a1}),
which includes the known cases $\nu=5$ and
$\nu=3$ $^{\m{\s{5}}}$, but extends the results to
arbitrary $\nu$ $(\geq 2)$. These transformation
brackets are particularly useful if one wants to evaluate
analytically certain quantities of physical interest,
in particular matrix elements of operators, such as the
electromagnetic transition operators, as discussed at
length in Ref.~6. In this article, the general result
for arbitrary $\nu \geq 2$ will be presented.

\section{SPECTRUM GENERATING ALGEBRA}

      As mentioned above, many applications of the method 
of spectrum generating algebras (SGA) for bound-state problems
in $\nu$ dimensions, have made use of the unitary algebra 
$U(\nu+1)$. By generalizing the well known
cases of $\nu=5$ $^{\m{\s{2}}}$ and $\nu=3$ $^{\m{\s{3}}}$,
we introduce
a realization of $U(\nu+1)$ in terms of $\nu+1$ boson operators,
divided into a set of $\nu$ operators, $b^{\dagger}_j$
($j=1,\ldots,\nu$), which transform as the fundamental
representation of $U(\nu)$ and an additional boson operator,
$b^{\dagger}_0=s^{\dagger}$, which transforms as a scalar
under $U(\nu)$ . The $\nu+1$ boson operators, $b^{\dagger}_j$
($j=0,\ldots,\nu$), span the $(\nu+1)$-dimensional space
of $U(\nu+1)$. The elements of $U(\nu+1)$ can be written
as the bilinear products
\ba\label{a2}
\begin{array}{ccc}
{\cal G} \, \equiv \, U(\nu+1) ~~:~~  & G_{jk}\;=\;b^{\dagger}_j b_{k}
& \hspace{1cm} (j,k=0,1,\ldots,\nu)~.
\end{array}
\ea
The states constructed by applying the boson creation operators
to a vacuum state
\ba\label{a3}
    {\cal B}: ~~~~~ \frac{1}{{\cal N}}  \, (b^{\dagger}_j)^{n_j} \,
(b^{\dagger}_{k})^{n_{k}} \ldots \, |0\rangle ~,
\ea
(where ${\cal N}$ is a normalization constant) transform
as the symmetric representation $[N]$ of
$U(\nu+1)$, where $N$ is the total number of bosons,
\ba\label{number}
\hat N \;=\; \sum_{j=0}^{\nu} \, \hat n_j
\;=\; \sum_{j=0}^{\nu} \, b^{\dagger}_j b_j~.
\ea
In the algebraic approach to bound state problems,
the Hamiltonian (and other) operators are expressed
as functions of the elements of $U(\nu+1)$, {\it i.e.}
they are in the enveloping algebra of $U(\nu+1)$,
and the basis states are the $[N]$ irreps of
$U(\nu+1)$.

      In this article we discuss
(i) the explicit construction of the basis states
in terms of bosons operators for $U(\nu+1) \supset
 U(\nu) \supset SO(\nu)$ and $U(\nu+1) \supset
SO(\nu+1) \supset SO(\nu)$, and
(ii) the transformation brackets relating the
basis states in the two chains.

\subsection{The chain $U(\nu+1) \supset U(\nu) \supset SO(\nu)$}

First we consider the chain 
\ba\label{chain1}
U(\nu+1) \supset U(\nu) \supset SO(\nu)~.
\ea
The basis states of this chain are denoted by
$| [N],n,\tau \rangle$, where $N$ is the total
number of bosons with $j =0,1,\ldots,\nu$ ,
describing the irreps $[N] \equiv [N,0,\ldots,0]$ of
$U(\nu+1)$, $n$ is the number of bosons with $j=1,\ldots,\nu$,
describing the irreps $[n] \equiv [n,0,\ldots,0]$ of $U(\nu)$
and $\tau$ is the quantum
number (boson seniority), describing the irreps
$(\tau) \equiv (\tau,0,\ldots,0)$ of $SO(\nu)$.
The branching rules are:
\ba\label{basis1}
\begin{array}{ll}
n \;=\; 0,1,\ldots,N~, & \\
\tau \;=\; n,n-2,\ldots,1 \m{ or } 0
& \hspace{1cm} (n \m{ odd or even, and } \nu > 2)~, \\
\tau \;=\; -n,-n+2,\ldots,n & \hspace{1cm} (\nu=2)~.
\end{array}
\ea
The branching of irreps of $SO(\nu)$ into
irreps of (eventual) subalgebras of $SO(\nu)$, is
of no interest for the present problem and
will not be discussed.

      The elements (generators) of $U(\nu)$ and
$SO(\nu)$ can be written as
\ba\label{gen1}
\begin{array}{rll}
U(\nu) ~~: ~~ & G_{jk}\;=\;b^{\dagger}_j b_{k}
& \hspace{1cm} (j,k=1,\ldots,\nu)~, \\
SO(\nu)~~: ~~ & L_{jk}\;=\;i \, (b^{\dagger}_j b_{k}-
b^{\dagger}_k b_{j}) & \hspace{1cm} (j<k \m{ and } j,k=1,\ldots,\nu)~.
\end{array}
\ea
The basis states of the chain (\ref{chain1}) can be written
in a compact form as
\ba\label{wf1}
| [N],n,\tau \rangle &=& \frac{1}{\sqrt{(N-n)!}} \,
(s^{\dagger})^{N-n} \, | [n],n,\tau \rangle ~,
\nonumber\\
| [n],n,\tau \rangle &=&
B_{n\tau} \, ( I_{\nu}^{\dagger} )^{(n-\tau)/2} \,
| [\tau],\tau,\tau \rangle ~.
\ea
The normalization coefficient $B_{n\tau}$ can be
derived by making use of the $SU(1,1)$ algebra, given 
in Eqs.~(\ref{gensu11}) and~(\ref{su11}) of Appendix A, 
\ba
B_{n\tau} &=& (-1)^{(n-\tau)/2} \,
\sqrt{ \frac{(2\tau+\nu-2)!!}{(n+\tau+\nu-2)!!(n-\tau)!!} } ~.
\label{bnt}
\ea
The operator $I^{\dagger}_{\nu}$ denotes the pair creation operator
in $\nu$ dimensions
\ba\label{pair1}
I^{\dagger}_{\nu} &=&
\sum_{j=1}^{\nu} \, b^{\dagger}_{j} b^{\dagger}_{j} ~,
\ea
and commutes with the generators of $SO(\nu)$,
$[I^{\dagger}_{\nu},L_{jk}]=0$~.
The operator $s^{\dagger}$ $(=b^{\dagger}_{0})$ has
been used in (\ref{wf1}) to make it conform with the standard
notation used in the literature.

\subsection{The chain $U(\nu+1) \supset SO(\nu+1) \supset SO(\nu)$}

Next we consider the chain 
\ba\label{chain2}
U(\nu+1) \supset SO(\nu+1) \supset SO(\nu) ~.
\ea
The basis states of this chain are denoted by
$| [N],\sigma,\tau \rangle$, where $[N]$ and $\tau$ label as
before the symmetric irreps of $U(\nu+1)$ and $SO(\nu)$, while
$(\sigma) \equiv (\sigma,0,\ldots,0)$ labels the
symmetric representation of $SO(\nu+1)$.
The branching rules are:
\ba\label{basis2}
\begin{array}{ll}
\sigma \;=\; N,N-2,\ldots,1 \m{ or } 0
& \hspace{1cm} (N \m{odd or even})~, \\
\tau \;=\; 0,1,\ldots,\sigma & \hspace{1cm} (\nu>2)~, \\
\tau \;=\; -\sigma,-\sigma+1, \ldots,\sigma & \hspace{1cm} (\nu=2)~.
\end{array}
\ea
The generators of $SO(\nu+1)$ and $SO(\nu)$ can be
written as
\ba\label{gen2}
\begin{array}{rll}
SO(\nu+1)~~: ~~ & L_{jk}\;=\;i \, (b^{\dagger}_j b_{k}-
b^{\dagger}_k b_{j}) & \hspace{1cm} (j<k \m{ and } j,k=0,\ldots,\nu)~, \\
SO(\nu)  ~~: ~~ & L_{jk}\;=\;i \, (b^{\dagger}_j b_{k}-
b^{\dagger}_k b_{j}) & \hspace{1cm} (j<k \m{ and } j,k=1,\ldots,\nu)~.
\end{array}
\ea
It is customary to separate the generators of $SO(\nu+1)$
into two pieces
\ba\label{gen3}
\begin{array}{ll}
L_{jk} \;=\; i \, (b^{\dagger}_j b_{k}-b^{\dagger}_k b_{j})
& \hspace{1cm} (j<k \m{ and } j,k=1,\ldots,\nu)~, \\
D_{j} \;=\; i \, (b^{\dagger}_0 b_j-b^{\dagger}_j b_0) \;=\;
i \, (s^{\dagger} b_j - b^{\dagger}_j s)
& \hspace{1cm} (j=1,\ldots,\nu)~.
\end{array}
\ea
Using the same $SU(1,1)$ algebra as discussed in Appendix A, but with 
the sum in (\ref{gensu11}) extending from $j=0$ to $j=\nu$, 
the basis states of the chain (\ref{chain2}) can be written as
\ba\label{wf2}
| [N],\sigma,\tau \rangle
&=& A_{N\sigma} \, ( I^{\dagger}_{\nu+1} )^{(N-\sigma)/2} \,
| [\sigma],\sigma,\tau \rangle ~.
\ea
Here $A_{N\sigma}$ is a normalization coefficient
\ba
A_{N\sigma} &=& (-1)^{(N-\sigma)/2} \,
\sqrt{ \frac{(2\sigma+\nu-1)!!}{(N+\sigma+\nu-1)!!(N-\sigma)!!}} ~,
\label{ans}
\ea
and $I^{\dagger}_{\nu+1}$ represents the pair creation operator in
$\nu+1$ dimensions
\ba\label{pair2}
I^{\dagger}_{\nu+1} &=&
\sum_{j=0}^{\nu} \, b^{\dagger}_{j} b^{\dagger}_{j}
\;=\; s^{\dagger} s^{\dagger} +
I^{\dagger}_{\nu} ~.
\ea
This pair creation operator commutes with the generators of $SO(\nu+1)$,
$[I^{\dagger}_{\nu+1},L_{jk}]=[I^{\dagger}_{\nu+1},D_j]=0$~.
In Appendix~B we show that
the states $| [\sigma],\sigma,\tau \rangle $ can be written as
\ba\label{wf3}
| [\sigma],\sigma,\tau \rangle &=&
\sum_{k=0}^{[(\sigma-\tau)/2]}
F_{k}(\sigma,\tau) \, ( s^{\dagger} )^{\sigma-\tau-2k} \,
( I^{\dagger}_{\nu+1} )^{k} \, | [\tau],\tau,\tau \rangle ~,
\ea
where the expansion coefficients are given by $^{\m{\s{7}}}$
\ba\label{fst}
F_{k}(\sigma,\tau) &=& \left[ \frac{(\sigma-\tau)!(2\tau+\nu-2)!!}
{(2\sigma+\nu-3)!!(\sigma+\tau+\nu-2)!} \right]^{1/2} \;
\left( -\frac{1}{2} \right)^{k}
\frac{(2\sigma+\nu-3-2k)!!}{(\sigma-\tau-2k)!k!} ~.
\ea
Another realization of
$SO(\nu+1)$ which is used frequently in physical applications, is
by the generators $\bar{D}_j=s^{\dagger} b_{j}+b^{\dagger}_{j} s$ with
$j=1,\ldots,\nu$ and
$L_{jk}=i(b^{\dagger}_j b_{k}-b^{\dagger}_k b_{j})$ with $j<k$ and
$j,k=1,\ldots,\nu$~.
The corresponding pair creation operator differs from
$I^{\dagger}_{\nu+1}$ in Eq.~(\ref{pair2}) by a relative sign
\ba
\bar{I}^{\dagger}_{\nu+1} &=&
s^{\dagger} s^{\dagger} - I^{\dagger}_{\nu} ~. \label{pair3}
\ea

\section{TRANSFORMATION BRACKETS}

      The transformation brackets between the two chain are
obtained by taking the overlap between the two sets
of basis states. Since both are written explicity in
terms of the states $| [\tau],\tau,\tau \rangle $, the
overlap is straightforward and yields the result
\ba\label{bra}
c^{\tau}_{n\sigma} &=& \langle [N],n,\tau \,
| \, [N],\sigma,\tau \rangle
\nonumber\\
&=& \sqrt{(N-n)!} \; \frac{A_{N\sigma}}{B_{n\tau}}
\sum_{k=k_0}^{[(\sigma-\tau)/2]} F_k(\sigma,\tau)
\left( \begin{array}{c} k+\frac{N-\sigma}{2} \\ \frac{n-\tau}{2}
\end{array} \right) ~,
\ea
with $k_0=\m{max}(0,\frac{1}{2}(n-\tau-N+\sigma))$.
For the second realization of the $SO(\nu+1)$ with the pair creation
operator of Eq.~(\ref{pair3}) the transformation
brackets have an additional sign $(-1)^{(n-\tau)/2}$~.
The transformation brackets of Eq.~(\ref{bra})
can be obtained by inserting the expressions
for the coefficients of Eqs.~(\ref{bnt},\ref{ans},\ref{fst})
\ba\label{bra1}
c^{\tau}_{n\sigma} &=& (-1)^{(N-\sigma-n+\tau)/2}
\left[ \frac{(N-n)!(n+\tau+\nu-2)!!(\sigma-\tau)!(2\sigma+\nu-1)}
{(N+\sigma+\nu-1)!!(N-\sigma)!!(\sigma+\tau+\nu-2)!(n-\tau)!!}
\right]^{1/2}
\nonumber\\
&& \sum_{k=k_0}^{[(\sigma-\tau)/2]} (-1)^k \,
\frac{(2\sigma+\nu-3-2k)!!(N-\sigma+2k)!!}
{(\sigma-\tau-2k)!(2k)!!(N-\sigma-n+\tau+2k)!!} ~,
\ea
or by introducing Pochhammer's symbol
$(a)_k=\Gamma(a+k)/\Gamma(a)$ as
\ba\label{bra2}
c^{\tau}_{n\sigma} &=&
\left( -\frac{1}{2} \right)^{(N-\sigma-n+\tau)/2} \,
(2\sigma+\nu-1)!!
\nonumber\\
&&\left[ \frac{(N-n)!(N-\sigma)!!(n+\tau+\nu-2)!!}
{(\sigma-\tau)!(n-\tau)!!(\sigma+\tau+\nu-2)!
(N+\sigma+\nu-1)!!(2\sigma+\nu-1)}
\right]^{1/2}
\nonumber\\
&& \sum_{k=k_0}^{[(\sigma-\tau)/2]} \frac{1}{k!} \,
\frac{((N-\sigma)/2+1)_k ((\tau-\sigma)/2)_k ((\tau-\sigma+1)/2)_k}
{((N-\sigma-n+\tau)/2+k)! (-\sigma-(\nu-3)/2)_k} ~.
\ea
Eqs.~(\ref{bra1}) and~(\ref{bra2}) reduce 
for $\nu=5$ and $\nu=3$ to the expressions
derived in Refs.~5, 8 and~9. 

For the lowest $SO(\nu+1)$ representation $\sigma=N$ the sum
appearing in the general expression for the transformation bracket
can be carried out explicitly to give
\ba
c^{\tau}_{nN} &=&
\left[ \frac{(N-\tau)!(N+\tau+\nu-2)!}
{(N-n)!(n+\tau+\nu-2)!!(n-\tau)!!(2N+\nu-3)!!} \right]^{1/2} ~,
\ea
in agreement with the results obtained previously $^{\m{\s{6}}}$ for 
$\nu=5$ and $\nu=3$.

Eqs.~(\ref{bra1}) and~(\ref{bra2}) conclude the derivations of the
transformation brackets for arbitrary $\nu$ $(\geq 2)$,
\ba
| [N],\sigma,\tau \rangle &=& \sum_{n} \, c^{\tau}_{n\sigma} \,
| [N],n,\tau \rangle ~.
\ea

\section{CONCLUSIONS}

      In this article, we have reported a closed expression for the
transformation brackets between the chains of Eq.~(\ref{a1}) for an
arbitrary number of dimensions $\nu$
$(\geq 2)$. These transformation brackets are
useful in a variety of problems which are being investigated at
the present time within the framework of the algebraic method.
For example, the case $\nu=2$ is of interest in the treatment
of bending vibrations of linear molecules, while the case
$\nu=9$ is of interest in the treatment of rotations and
vibrations of non-planar tetratomic molecules.

      The transformation brackets derived here can be used to
evaluate matrix elements of an operator $\hat T$
in the `deformed' chain
(of great physical interest) by a two-step process, {\it i.e.}
by first evaluating them in the `spherical' chain (which is a
relatively easy calculation) and subsequently transforming the
results to the `deformed' chain
\ba
\langle [N],\sigma^{\prime},\tau^{\prime} | \,
\hat T \, | [N],\sigma,\tau \rangle &=&
\sum_{n^{\prime},n} \, c^{\tau^{\prime}}_{n^{\prime}\sigma^{\prime}} \,
c^{\tau}_{n\sigma} \, \langle [N],n^{\prime},\tau^{\prime} | \,
\hat T \, | [N],n,\tau \rangle ~,
\ea
where the coefficients $c$ are the transformation brackets
derived in this article.

\section*{ACKNOWLEDGEMENTS}
      This work was supported in part by I.N.F.N.,
by CONACyT, M\'exico under project 400340-5-3401E, DGAPA-UNAM under
project IN105194, and by D.O.E. Grant DE-FG02-91ER40608.

\appendix 
\section{THE REDUCTION $U(\nu) \supset SO(\nu)$}
\setcounter{equation}{0}

       The normalization coefficients $B_{n\tau}$ and
$A_{N\sigma}$ of Eqs.~(\ref{bnt}) and~(\ref{ans}) can be found
by making use of the properties of the $SU(1,1)$ `quasi-spin' 
algebra for a system of bosons $^{\m{\s{10,11}}}$.
Consider a system of bosons
in $\nu$ dimensions with the algebraic structure
\ba
U(\nu) \supset SO(\nu) ~.
\ea
The corresponding basis states are characterized by $|[n],\tau \rangle$~.
The operators
\ba
\hat Q_+ &=& \frac{1}{2} \sum_{j=1}^{\nu} \, b_j^{\dagger} b_j^{\dagger} ~,
\nonumber\\
\hat Q_- &=& \frac{1}{2} \sum_{j=1}^{\nu} \, b_j b_j ~,
\nonumber\\
\hat Q_0 &=& \frac{1}{4} \sum_{j=1}^{\nu} \,
( b_j^{\dagger} b_j + b_j b_j^{\dagger} )
\;=\; \frac{1}{2}(\hat n + \frac{1}{2}\nu) ~, \label{gensu11}
\ea
satisfy the commutation relations
\ba
\, [\hat Q_+,\hat Q_-] &=& -2 \, \hat Q_0 ~,
\nonumber\\
\, [\hat Q_0,\hat Q_{\pm}] &=& \pm \, \hat Q_{\pm} ~, \label{su11}
\ea
of the $SU(1,1)$ algebra.
The basis states are labeled by $|q,q_0\rangle$.
The generators of the $SU(1,1)$ algebra defined in (\ref{gensu11}) 
commute with the generators of $SO(\nu)$, $L_{jk}$ of Eq.~(\ref{gen1}).
The relation between the two sets of basis states $|q,q_0\rangle$ and
$|[n],\tau\rangle$ can be found by using the generators 
of (\ref{gensu11}).
First, $\hat Q_0$ is diagonal in the basis states
\ba
\hat Q_0 \, |q,q_0 \rangle &=& q_0 \, |q,q_0 \rangle ~,
\nonumber\\
\hat Q_0 \, |[n],\tau \rangle &=& \frac{1}{2}(n+\frac{1}{2}\nu) \,
|[n],\tau \rangle ~,
\ea
and hence $q_0=\frac{1}{2}(n+\frac{1}{2}\nu)$~.
Furthermore, $\hat Q_-$ annihilates the highest-weight state
\ba
\hat Q_- \, |q=q_0,q_0 \rangle &=&
\hat Q_- \, |[n=\tau],\tau \rangle \;=\; 0 ~,
\ea
which gives $q=\frac{1}{2}(\tau+\frac{1}{2}\nu)$~.
The basis states can be expanded as
\ba
|q,q_0\rangle &=& A_{qq_0} \, (\hat Q_+)^{q_0-q} \, |q,q_0=q\rangle ~,
\nonumber\\
A_{qq_0} &=& (-1)^{q_0-q} \,
\sqrt{ \frac{(2q-1)!}{(q_0-q)!(q_0+q-1)!} } ~, \label{qq0}
\ea
or alternatively as
\ba
|[n],\tau \rangle &=& B_{n\tau} \, (I^{\dagger}_{\nu})^{(n-\tau)/2}
\, |[\tau],\tau \rangle ~,
\nonumber\\
B_{n\tau} &=& (-1)^{(n-\tau)/2} \,
\sqrt{ \frac{(2\tau+\nu-2)!!}{(n+\tau+\nu-2)!!(n-\tau)!!} } ~.
\ea
The choice of phase in (\ref{qq0}) is conventional.

\section{THE REDUCTION $SO(\nu+1) \supset SO(\nu)$}
\setcounter{equation}{0}

      The $SU(1,1)$ algebra of (\ref{gensu11}) can be used for the
reductions $U(\nu) \supset SO(\nu)$ and $U(\nu+1) \supset SO(\nu+1)$.
For the chain (\ref{chain2}) we need the further reduction
$SO(\nu+1) \supset SO(\nu)$. This is by far more complex.
We use here Ref.~12  and a generalization of the method
discussed on pages 152-157 of Ref.~5. The
defining equations are
\ba
\hat N \, |[N],\sigma,\tau \rangle &=& N \, |[N],\sigma,\tau \rangle ~,
\nonumber\\
\hat C_{SO(\nu+1)} \, |[N],\sigma,\tau \rangle &=&
\sigma(\sigma+\nu-1) \, |[N],\sigma,\tau \rangle ~,
\nonumber\\
\hat C_{SO(\nu)} \, |[N],\sigma,\tau \rangle &=&
\tau(\tau+\nu-2) \, |[N],\sigma,\tau \rangle ~. \label{eigen}
\ea
The notation for the states is the same as in Section II.A.
$\hat C_G$ represents the quadratic Casimir invariant of $G$.
These equations can be expressed in terms of a set of separable
differential equations by introducing hyperspherical coordinates
\ba
(x_1,\ldots,x_{\nu+1}) &\rightarrow&
(r,\phi,\theta_{\nu-1},\ldots,\theta_1) ~,
\ea
by
\ba
x_1 &=& r \, \sin \phi \, \sin \theta_{\nu-1} \, \cdots \, \sin \theta_2
\, \cos \theta_1 ~,
\nonumber\\
x_2 &=& r \, \sin \phi \, \sin \theta_{\nu-1} \, \cdots \, \sin \theta_2
\, \sin \theta_1 ~,
\nonumber\\
x_3 &=& r \, \sin \phi \, \sin \theta_{\nu-1} \, \cdots \, \cos \theta_2 ~,
\nonumber\\
& \vdots &
\nonumber\\
x_{\nu} &=& r \, \sin \phi \, \cos \theta_{\nu-1} ~,
\nonumber\\
x_{\nu+1} &=& r \, \cos \phi ~,
\ea
with $0 \leq r < \infty$ and
$0 \leq \phi,\theta_{\nu-1},\ldots,\theta_2 < \pi$ and
$0 \leq \theta_1 < 2\pi$~.
The volume element is given by
\ba
dx_1 \, \cdots \, dx_{\nu+1} &=& r^{\nu} \, (\sin \phi)^{\nu-1} \,
(\sin \theta_{\nu-1})^{\nu-2} \, \cdots \, \sin \theta_2 \,
dr \, d\phi \, d\theta_{\nu-1} \, \cdots \, d\theta_2 \, d\theta_1 ~.
\ea
The Casimir invariants can be obtained from the Laplacian in
$\nu+1$ dimensions and a recursion relation between the Casimir
operators of the orthogonal groups (see page 493 of Ref.~12)
\ba
\nabla^2_{\nu+1} &=& \frac{1}{r^{\nu}} \frac{\partial}{\partial r}
\left( r^{\nu} \frac{\partial}{\partial r} \right) - \frac{1}{r^2} \,
\hat C_{\nu+1}(\phi,\theta) ~,
\nonumber\\
\hat C_{\nu+1}(\phi,\theta) &=& - \frac{1}{(\sin \phi)^{\nu-1}}
\frac{\partial}{\partial \phi}
\left( (\sin \phi)^{\nu-1} \frac{\partial}{\partial \phi} \right)
+ \frac{1}{(\sin \phi)^2} \, \hat C_{\nu}(\theta) ~,
\ea
Here $(\theta)=(\theta_{\nu-1},\ldots,\theta_1)$.
The number operator expressed in hyperspherical coordinates is
given by
\ba
\hat N &=& \frac{1}{2} \left[ - \frac{1}{r^{\nu}}
\frac{\partial}{\partial r}
\left( r^{\nu} \frac{\partial}{\partial r} \right) + r^2
+ \frac{1}{r^2} \, \hat C_{\nu+1}(\phi,\theta) - (\nu+1) \right] ~.
\ea
The eigenvector equations can be solved by
separation of variables and have solutions in terms of products
of Laguerre and Gegenbauer polynomials
\ba
\psi_{N\sigma\tau\alpha} (r,\phi,\theta) &=& f_{N\sigma}(r) \,
g_{\sigma \tau}(\phi) \, \Phi_{\tau\alpha}(\theta) ~,
\nonumber\\
&=& A_{N\sigma\tau} \, r^{\sigma} \, \mbox{e}^{-r^2/2} \,
L_{(N-\sigma)/2}^{(2\sigma+\nu-1)/2}(r^2) \,
(\sin \phi)^{\tau} C^{(2\tau+\nu-1)/2}_{\sigma-\tau}(\cos \phi)
\, \Phi_{\tau\alpha}(\theta) ~,
\nonumber\\
\ea
with
\ba
A_{N\sigma\tau} &=& (-1)^{(N-\sigma)/2} (2\tau+\nu-3)!! \left[
\frac{2^{(2\sigma+\nu+1)/2}(2\sigma+\nu-1)(N-\sigma)!!(\sigma-\tau)!}
{\pi(N+\sigma+\nu-1)!!(\sigma+\tau+\nu-2)!} \right]^{1/2} ~.
\nonumber\\
\ea
Next we apply Dragt's theorem to the highest weigth state with $N=\sigma$
\ba
\psi_{\sigma\sigma\tau\alpha} (r,\phi,\theta) &=&
\frac{A_{\sigma\sigma\tau}}{A_{\tau\tau\tau}} \, r^{\sigma-\tau} \,
C^{(2\tau+\nu-1)/2}_{\sigma-\tau}(\cos \phi) \,
\psi_{\tau\tau\tau\alpha} (r,\phi,\theta) ~,
\ea
by replacing
\ba
r &\rightarrow& ( I^{\dagger}_{\nu+1}/2 )^{1/2} ~,
\nonumber\\
\cos \phi &\rightarrow& t^{\dagger}
\;=\; s^{\dagger}/(I^{\dagger}_{\nu+1})^{1/2} ~,
\ea
to obtain
\ba
|[\sigma],\sigma,\tau \rangle &=&
\frac{A_{\sigma\sigma\tau}}{A_{\tau\tau\tau}} \,
( I^{\dagger}_{\nu+1}/2 )^{(\sigma-\tau)/2} \,
C^{(2\tau+\nu-1)/2}_{\sigma-\tau}(t^{\dagger}) \,
|[\tau],\tau,\tau \rangle ~,
\nonumber\\
&=& \left[ \frac{(\sigma-\tau)!(2\tau+\nu-2)!!}
{(2\sigma+\nu-3)!!(\sigma+\tau+\nu-2)!} \right]^{1/2}
\sum_{k=0}^{[(\sigma-\tau)/2]} \left( -\frac{1}{2} \right)^k
\nonumber\\
&& \frac{(2\sigma+\nu-3-2k)!!}{k!(\sigma-\tau-2k)!} \,
( s^{\dagger} )^{\sigma-\tau-2k} \, ( I^{\dagger}_{\nu+1} )^k \,
|[\tau],\tau,\tau \rangle ~. \label{wf}
\ea
By comparing (\ref{wf}) with (\ref{wf3}), one finds the expression of
Eq.~(\ref{fst}) for the expansion coefficients $F_{k}(\sigma,\tau)$.
$$ $$
\scriptsize

$^{\m{\s{1}}}$ F. Iachello, in Lie Algebras, Cohomologies
and New Applications of Quantum Mechanics, Contemporary Mathematics, AMS,
N. Kamran and P. Olver, eds. Vol.160, p.161-171 (1994).

$^{\m{\s{2}}}$ F. Iachello and A. Arima, The Interacting Boson Model,
Cambridge University Press, Cambridge (1987).

$^{\m{\s{3}}}$  F. Iachello and R.D. Levine, Algebraic
Theory of Molecules, Oxford University Press, Oxford (1995).

$^{\m{\s{4}}}$  R. Bijker, F. Iachello and A. Leviatan,
Ann. Phys. (N.Y.) {\bf 236}, 69 (1994);
R. Bijker and A. Leviatan, Revista Mexicana de F\ii sica {\bf 39},
Suplemento 2, 7 (1993).

$^{\m{\s{5}}}$
A. Frank and P. van Isacker,
Algebraic Methods in Molecular and Nuclear Physics,
Wiley-Interscience (1994).

$^{\m{\s{6}}}$
A. Arima and F. Iachello,
Ann. Phys. (N.Y.) {\bf 123}, 468 (1979).

$^{\m{\s{7}}}$
R. Bijker and J.N. Ginocchio,
Phys. Rev. {\bf C45}, 3030 (1992).

$^{\m{\s{8}}}$
O. Casta\~nos, E. Chac\'on, A. Frank and M. Moshinsky,
J. Math. Phys. {\bf 20}, 35 (1979).

$^{\m{\s{9}}}$
A. Frank and R. Lemus,
J. Chem. Phys. {\bf 84}, 2698 (1986); Ann. Phys. (N.Y.) {\bf 206},
122 (1991).

$^{\m{\s{10}}}$
H.Ui, Ann. Phys. (N.Y.) {\bf 49}, 69 (1968).

$^{\m{\s{11}}}$
A. Arima and F. Iachello,
Ann. Phys. (N.Y.) {\bf 99}, 253 (1976).

$^{\m{\s{12}}}$
N. Ya. Vilenkin, Special Functions and the Theory of Group
Representations, AMS Translations, Providence, R.I. (1968),
Volume 22.

\normalsize

\end{document}